\documentclass[lettersize,journal]{IEEEtran}
\usepackage{epsfig,rotating,setspace,latexsym,amsmath,epsf,amssymb,amsfonts,bm,theorem,cite,enumerate,longtable,accents,float,physics}
\usepackage{colortbl}
\usepackage{array}
\usepackage{algorithm,algorithmic,graphicx,epsf,authblk,epstopdf,url,xcolor, soul,multirow,bbm,subcaption}

\usepackage{mathtools,comment}
\usepackage[center]{qtree}
\usepackage{tree-dvips}
\usepackage[linguistics]{forest }
\usepackage{diagbox}

\usepackage[utf8]{inputenc} 
\usepackage[T1]{fontenc}
\usepackage{ifthen}
\usepackage{cite}

\newtheorem{theorem}{Theorem}

\newtheorem{definition}{Definition}

\newtheorem{remark}{Remark}

\newenvironment{Proof}[1]{\medskip\par\noindent{\bf Proof:\,}\,#1}{{\mbox{\,$\blacksquare$}\par}}

\newcommand{\cq}{{\mathcal{Q}}}
\newcommand{\cR}{{\mathcal{R}}}

\newcommand{\cs}{{\mathcal{S}}}

\DeclareMathOperator{\diag}{diag}

\allowdisplaybreaks
\date{}

\begin{document}

\title{Quantum Private Distributed Matrix Multiplication: Extending the Classical Codes and Limitations}

\author{Mohamed Nomeir \quad Alptug Aytekin \quad Lei Hu \quad   Sennur Ulukus \\
    \normalsize Department of Electrical and Computer Engineering\\
    \normalsize University of Maryland, College Park, MD 20742 \\
    \normalsize \emph{mnomeir@umd.edu} \quad \emph{aaytekin@umd.edu} \quad \emph{leihu@umd.edu} \quad \emph{ulukus@umd.edu}}

\maketitle

\begin{abstract}
    In this paper, we explore how quantum resources can be used to increase the rate of private distributed matrix multiplication (PDMM). In PDMM, a user who has two high-dimensional matrices, $A$ and $B$, and lacks the computational capabilities to apply matrix multiplication locally, divides the matrices $A$ and $B$ into $K$ and $L$ sub-blocks, respectively. Then, the user sends them to $N$ servers to apply the required multiplication \emph{privately}, i.e., any $T$ colluding servers cannot get any information about the user's matrices. The goal is to reduce the number of servers needed to perform the required matrix multiplication, thereby decreasing the communication cost. In the quantum setting, we allow the servers to share an entangled state, amongst each other only, and respond back to the user over separate quantum channels. Upon receiving the quantum dits, the user should be able to obtain the required matrix multiplication by applying the appropriate measurements. There are two main regimes in the PDMM literature (assuming $K \geq L$, without loss of generality): The first regime is when the privacy parameter $T$ is high compared to $K$ or $L$, i.e., $T>K\geq L$ or $K \geq T \geq L$, called the high-privacy regime.  In the second regime, the low-privacy regime, the privacy parameter $T$ is less than $K$ and $L$, i.e., $K\geq L > T$. 
    
    First, in the high-privacy regime, the state-of-the-art classical code is called the gap additive secure polynomial (GASP) code. We define a feasibility requirement in the quantum setting for the GASP code such that the highest performance is achieved when the requirement is satisfied. Thus, super-dense coding gain is achieved when the feasibility condition is satisfied. Then, we analyze the relationship between $K$, $L$ and $T$ for the feasibility condition to be satisfied. We show that when $T \geq KL-K+1$, the feasibility condition is always satisfied and the GASP code can be extended to the quantum version. In the case of $T < KL-K+1$, the feasibility can still be satisfied. To further examine this behavior, we numerically study how the minimum privacy requirement depends on the matrix dimensions and provide a quadratic estimate for this relation. The results suggest that feasibility can be achieved when $T \sim 0.5 KL$. Second, in the low-privacy regime, the recently developed cyclic-addition degree tables (CAT) and discretely optimized GASP (DOG) codes are among the most efficient known classical constructions for PDMM. We show that the feasibility condition developed for GASP can be adopted for both CAT and DOG codes as well, thus unifying the feasibility framework for multiple classical PDMM coding schemes. 
\end{abstract}


\section{Introduction}
Matrix multiplication (MM) is a key component in all machine learning (ML), artificial intelligence (AI), and signal processing (SP) tasks \cite{mm_1, mm_2, mm_3}. Due to the high dimensionality of the data, matrix multiplication often involves high-dimensional matrices that cannot be computed or processed using a single local computer. The idea of relaying these operations to more computationally capable entities, which we call servers, was introduced to address this limitation in the computer science community \cite{cs_1, cs_2, cs_3, cs_4, cs_5, cs_6}. The user who distributes the matrices to the servers seeks to preserve the privacy of the data. Thus, the main goal is to hide as much information about these matrices as possible from the computing servers, while ensuring low computation complexity and high download rates. Recently, the problem has been adopted into the communications and information theory communities, and it was first formalized by \cite{ravi_1}. In \cite{ravi_1}, the problem was formulated as a user who has two matrices, $A$ and $B$, for which the user does not have the capability to compute their product. The user sends a coded version of each matrix to ensure privacy, after slicing $A$ horizontally and  $B$ vertically, to $N$ servers to compute the multiplication of the coded versions and transmit the multiplication back to the user. 

There are two privacy settings adopted in \cite{ravi_1}. The first requires that the two matrices, $A$ and $B$, to be kept \emph{information-theoretically} private from any $T$ curious but honest servers that share their coded versions with each other. The main performance metric is the download rate, i.e., the number of useful multiplications over the number of downloaded symbols. The second setting in \cite{ravi_1} analyzes the case when $A$ is kept private but $B$ is public knowledge. The latter setting is not considered in this paper, and we only consider the first case with both $A$ and $B$ being private. The approach for slicing $A$ horizontally into $K$ sub-matrices and $B$ vertically into $L$ sub-matrices and transmitting them to the curious servers appears to increase the total computational requirements compared to the single computer at the user side due to the privacy requirement. This concern was echoed in \cite{jafar_SDBMM}. However, in \cite{response_to_jafar}, it was shown that indeed \emph{this is not the case}, and computational complexity can be optimized if the correct partitioning is used. Finally, the act of slicing $A$ horizontally and $B$ vertically and requiring $AB$ computed is called outer product partitioning (OPP), and it is the main focus in this paper. If the required product is $BA$, then it is called inner product partitioning (IPP); see e.g., \cite{inner_1, inner_2, inner_3}. Other works have considered more general frameworks for both the IPP and OPP, e.g., in \cite{general_1, general_2, general_3, general_4, general_5, general_6}.

Since the OPP problem is a critical step in a wide range of applications such as GPU acceleration, tensor contractions and deep learning \cite{app_1,app_2,app_3}, many works have studied the OPP problem defined in \cite{ravi_1} in different contexts. In \cite{bivariate,multivariate}, bivariate and multivariate polynomials are used to encode the submatrices. Specifically, bivariate polynomials are sufficient to be used in the case of homogeneous upload cost thresholds, and multivariate polynomials are suitable for the case of stragglers. In \cite{precomputation}, the case where the user has some information about the multiplication of the noise matrices is analyzed. In \cite{AG_OPP}, the OPP problem is studied in the context of algebraic geometric codes. In \cite{approx_opp}, successive cancellation is used to obtain an approximate result for the computation of $AB$ that becomes more accurate with time. 

A widely adopted technique for OPP is called gap additive secure polynomial (GASP) method, developed in \cite{Gasp_1}. GASP utilizes polynomial codes and selects polynomials to minimize the required number of servers. Until recently, GASP was regarded as the state-of-the-art for the PDMM OPP across all privacy regimes. However, in a recent work, \cite{cat-dog} introduced three novel code constructions. The three codes are modified versions of GASP: 1) optimized GASP, which provides better rates than GASP in general at the expense of requiring high computational power to find the optimal number of servers, 2) the cyclic-addition degree tables (CAT), and 3) the discretely optimized GASP (DOG). Both CAT and DOG outperform GASP or optimized GASP in low-privacy regimes, while optimized GASP consistently outperforms GASP. The first reason why we adopt GASP over optimized GASP is the computational complexity, especially when the sub-packetization of the matrices and the privacy requirement are high, i.e., when $K$, $L$ and $T$ are large, since the optimal value of the number of servers is found by trial and error over two free parameters (related to $K$ and $T$) as opposed to one free parameter for GASP. The second reason we adopt GASP over optimized GASP is that the normal GASP is more likely to satisfy the required feasibility constraint we developed to extend to the quantum version, compared to the optimized GASP, according to our trials. When the feasibility requirement is satisfied, the classical code can be extended to the quantum code, achieving double-the-rate of its classical counterpart.\footnote{{In this work, the terms security and privacy are used interchangably. We use both to mean the same thing, since in this setting, keeping the matrices secure means keeping them private from any colluding servers.}}

Concurrently, quantum communications has gained significant attention recently due to its potential in increasing the fundamental limits of communication compared to classical communications. It was shown in \cite{super_dense} that pre-entanglement between the transmitter and the receiver can double the communication rate, i.e., a single qubit can carry two bits of useful information, which was coined as \emph{super-dense coding}. The setting where $N$ entangled transmitters communicate over a noiseless quantum channel to a receiver was re-formalized in \cite{nsumbox}. In that setting, at most $N$ useful bits can be received by the receiver. In normal communication scenarios with no other requirement, i.e., privacy or security requirements or specific summations from multi-server systems, this restriction implies that no gain can be achieved in using the quantum resources compared to the classical ones. However, quantum abilities showed great promise in both the quantum private information retrieval (QPIR) setting and the quantum computation setting, where super-dense coding gain was achieved, as shown in \cite{our_quantum_first, our_journal, Byzantine_journal, byzantine_1, our_symm_byz_quantum_journal, jafar_quantum_unresponsive, lu_product,  yao_capacity_MAC, yao_inverted}. The setting in \cite{nsumbox} and the corresponding formulation, using specific encoding and decoding, was coined as the \emph{$N$-sum box abstraction}. 

In this paper, we aim to develop a framework for PDMM that extends the classical codes to the quantum setting when the transmitters are entangled and share separate quantum channels with the user. We aim to improve the communication efficiency of PDMM in the multiple-to-one quantum network. The main contributions of our work are summarized as:
\begin{itemize}
    \item We formulate the PDMM problem in the quantum communication framework and study the problem in two different regimes, the low privacy regime and the high privacy regime, respectively.
    \item For the high privacy regime, we study the extension of the GASP codes to the quantum setting. We provide a feasibility condition, such that, when satisfied, the maximum communication efficiency can be achieved, i.e., we maintain the same number of servers while doubling the download rate compared to its classical counterpart, which achieves super-dense coding gain.
    \item For the non-degenerate case where $K,~L\geq 2$, we prove that when $T \geq KL-K+1$, consequently when $T> KL$, the GASP code can be extended to its quantum counterpart. When $T < KL-K+1$, our numerical results suggest that if $T_{\rm min} \sim 0.5 KL$ and $T \geq T_{\rm min}$ the feasibility condition is satisfied by the GASP code.
    \item For the low privacy regime, we utilize the aforementioned three codes in \cite{cat-dog}, and show that the feasibility condition defined for GASP can be utilized for them as well, thus unifying the approach. Specifically, we focus on CAT and DOG codes since they provide higher rates compared to the GASP codes and the optimized GASP codes in low privacy regimes. We provide an explicit example where the CAT code surpasses all the aforementioned codes and satisfies the feasibility constraint, and thus, we achieve double-the-rate compared to the highest possible classical rate to date.
\end{itemize}

Next, we summarize our results and findings: In this paper, we extend the GASP codes to their quantum counterpart via a feasibility condition that we developed. When the condition is satisfied, we can apply our proposed scheme to double the download rate. Also, we show that the feasibility condition is always satisfied when $T \sim 0.5KL$. Furthermore, we focus on the classical codes that are efficient in the low privacy regime. We extend the modified GASP, CAT, and DOG codes using the same feasibility condition, thus unifying the framework. We note that, when the privacy requirement is low, it is hard to find a case when CAT and DOG surpass the GASP code and satisfy the feasibility condition at the same time. However, we show that when $K=L=T=2$, the CAT code surpasses all the classical PDMM OPP codes and satisfies the feasibility requirement. 

The paper is organized as follows: Section \ref{review} reviews important topics related to our work, namely, generalized Reed-Solomon (GRS) and dual GRS codes, quantum information theory, and the $N$-sum box abstraction. In addition, we provide some important minor extension results that are needed in our paper. Section \ref{sec_prob_form} formulates the PDMM problem in the classical and quantum settings. Section \ref{sec_gasp_all} focuses on the GASP codes for the high privacy regime. In Section \ref{sec_gasp_all}, we review the classical GASP code, then we provide a condition such that, if satisfied, the GASP code can be extended to the quantum version and achieve super-dense coding gain. Then, we prove that the aforementioned condition, the feasibility condition, is satisfied whenever $T \geq KL-K+1$. We end Section \ref{sec_gasp_all} with numerical analysis to give us an intuitive understanding of the applicability of the feasibility condition.  In Section \ref{sec_more_pdmm_low_privacy}, we review the classical PDMM codes that achieve better rates compared to the GASP code in the low privacy regime and use the same condition to extend them to the quantum version. In addition, we provide a detailed example in this case to show that the super-dense coding gain can be achieved if the aforementioned condition is satisfied.

\textit{Notation:} Prior to delving into the next sections, we explicitly list the general notations used, except stated otherwise, for the paper. We denote the natural numbers as $\mathbb{N} = \{0,1,2, \ldots\}$. Let $\alpha \in \mathbb{N}^{K+T}$ and $\beta \in \mathbb{N}^{L+T}$, then we denote 
\begin{align}
    \alpha \oplus \beta 
    = \begin{bmatrix}
      \alpha_1 + \beta_1& \alpha_1+\beta_2& \ldots & \alpha_1+\beta_{L+T}\\
      \alpha_2 + \beta_1& \alpha_2+\beta_2& \ldots & \alpha_2+\beta_{L+T}\\
      \vdots & \vdots& \ddots & \vdots\\
      \alpha_{K+T} + \beta_1& \alpha_{K+T}+\beta_2& \ldots & \alpha_{K+T}+\beta_{L+T}\\
    \end{bmatrix}.
\end{align}
If $A \in \mathbb{R}^{m \times n}$, then let $Set(A)$ denote the set with the elements of the matrix $A$. We use the superscript $t$ for the transpose operator, with computational basis as the main basis, and $\dagger$ is the conjugate transpose operator. $[n]$ denotes the set $\{1, \ldots, n\}$, and $[a:b]$ denotes the set $\{a,a+1, \ldots, b\}$. For a matrix $A \in \mathbb{R}^{m \times n}$, $A(:,[a:b])$ defines a sub-block of the aforementioned matrix with all the rows included and the columns chosen are those indexed by the set $[a:b]$. All vectors are column vectors except stated otherwise. All symbols that denote sets are written using a calligraphic font. If $\cs$ is a set, then $\cs(m)$ denotes the $m$th element in the set after a predefined indexing. If $a$ is any constant, then $a^{\cs} = [a^{\cs(1)}, \ldots, a^{\cs(|\cs|)}] $. If $K$ is a constant and $a=[a_1, \ldots, a_n]$, then, $K + a = K \mathbf{1}^t + a$, where $\mathbf{1}$ is the all ones vector with the required length. We use $\mathbbm{1}_{\cs}$ as an indicator random variable for the occurrence of the event $\cs$. $I_N$ is the identity matrix of size $N$. Uppercase letters are used for matrices, and lowercase letters are used for vectors. Constants are written in upper or lower case fonts depending on the situation, and will be mentioned if not clear from the context. Finally, for the quantum systems and quantum operators, we use upper case with typewriter font, and for the associated quantum density matrix, as a special case of quantum operators, we use the Greek symbols $\sigma$ and $\rho$.

\section{Preliminaries} \label{review}
\subsection{Dual Codes}
Let $C$ be a linear code in $\mathbb{F}_q^n$, the dual linear code of $C$, $C^{\perp}$, is defined as 
\begin{align}
    C^{\perp} = \{x \in \mathbb{F}_q^n: x^tc=0, ~c \in C \}.
\end{align}
Since both are linear codes, they can be generated using generator matrices, $G_{C} \in \mathbb{F}_{q}^{k \times n}$, and $G_{C^{\perp}} \in \mathbb{F}_{q}^{n-k \times n}$, respectively. Then, they are connected with the relation
\begin{align}
    G_{C}G_{C^{\perp}}^t = 0.
\end{align}
An important code family in this work is the GRS code and its dual \cite{dual_grs_ref}. The $[n,k,d]_q$ GRS code is defined as follows.

\begin{definition}[GRS code]
    Let $F = \mathbb{F}_q$ be a field and choose $n$ distinct non-zero elements $\alpha_1, \ldots, \alpha_n$. Let $u_1, \ldots, u_n$ be non-zero elements in the field. Then, for any $0 \leq k \leq n$, the GRS is defined as
    \begin{align}
        \text{GRS}(\alpha, u) = & \{\left (u_1f(\alpha_1), u_2f(\alpha_2), \ldots, u_nf(\alpha_n)\right): \nonumber \\
        &\quad f(x) \in F[x]_k \},
    \end{align}
where $\alpha=\alpha_{[n]}$, $u = u_{[n]}$, and $F[x]_k$ is the polynomial ring with degree less than $k$. In addition, the corresponding canonical generator matrix (transposed) is given by
\begin{align}
    G_{\text{GRS}(\alpha, u)} = \begin{bmatrix}
        u_1& u_1\alpha_1&u_1\alpha_1^2 & \ldots& u_1 \alpha_1^{k-1}\\
        u_2& u_2\alpha_2&u_2\alpha_2^2 & \ldots& u_2 \alpha_2^{k-1}\\
        \vdots& \vdots&\vdots & \ddots& \vdots\\
        u_n& u_n\alpha_n&u_n\alpha_n^2 & \ldots& u_n \alpha_n^{k-1}\\
    \end{bmatrix}.
\end{align}
\end{definition}
For notational convenience, we henceforth refer to the transposed generator matrix simply as the generator matrix without loss of generality.

\begin{theorem}[Dual GRS]\label{thm_dual_grs}
    Given an $[n,k,d_1]_q$ GRS code parameters with $\alpha$ and $u$ as the generating elements, the dual code is also a $[n,n-k,d_2]_q$ GRS code with $\alpha$ and $v = v_{[n]}$ as the generating elements, where
    \begin{align}\label{dual_grs_relation}
        v_i = \frac{1}{u_i} \left( \prod_{j \in [n]\atop j \neq i }(\alpha_j-\alpha_i)\right)^{-1}.
    \end{align}
\end{theorem}

From Theorem \ref{thm_dual_grs}, the generator matrix for the dual GRS can be written as
\begin{align}
    G_{\text{GRS}(\alpha, v)} = \begin{bmatrix}
        v_1& v_1\alpha_1&v_1\alpha_1^2 & \ldots& v_1 \alpha_1^{n-k-1}\\
        v_2& v_2\alpha_2&v_2\alpha_2^2 & \ldots& v_2 \alpha_2^{n-k-1}\\
        \vdots& \vdots&\vdots & \ddots& \vdots\\
        v_n& v_n\alpha_n&v_n\alpha_n^2 & \ldots& v_n \alpha_n^{n-k-1}\\
    \end{bmatrix}.
\end{align}

\subsection{Dual of the Shifted GRS Code}
\begin{definition}[Shifted GRS]
    An $[n,k,d]_q$ GRS code is called $\ell$-shifted GRS if the generator matrix is defined as
    \begin{align}
        G_{\text{GRS}(\alpha, u)} = \begin{bmatrix}
        u_1\alpha_1^{\ell}& u_1\alpha_1^{\ell+1}&u_1\alpha_1^{\ell+2} & \ldots& u_1 \alpha_1^{\ell+k-1}\\
        u_2\alpha_2^{\ell}& u_2\alpha_2^{\ell+1}&u_2\alpha_2^{\ell+2} & \ldots& u_2 \alpha_2^{\ell+k-1}\\
        \vdots& \vdots&\vdots & \ddots& \vdots\\
        u_n\alpha_n^{\ell}& u_n\alpha_n^{\ell+1}&u_n\alpha_n^{\ell+2} & \ldots& u_n \alpha_n^{\ell+k-1}\\
        \end{bmatrix}.
    \end{align}
\end{definition}

\begin{theorem}
    For any $[n,k,d_1]_q$ $\ell_1$-shifted GRS code with $u$ and $\alpha$ as the generating elements, there exists an $[n,n-k,d_2]_q$ $\ell_2$-shifted GRS code with $v$ and $\alpha$ as its dual code if
    \begin{align}
        v_i = \frac{1}{u_i \alpha_i^{\ell_1+\ell_2}} \left( \prod_{j \in [n]\atop j \neq i }(\alpha_j-\alpha_i)\right)^{-1}.
    \end{align}
\end{theorem}

This result follows directly by substituting \eqref{dual_grs_relation} and plugging $u_i$ as $u_i\alpha_i^{\ell_1}$ and $v_i$ as $v_i \alpha_i^{\ell_2}$, respectively.

\subsection{Quantum Information Theory}
For completeness, we recall the following standard definitions from quantum information theory. Detailed explanation and properties can be found in standard references on quantum information processing, e.g., \cite{nielsen-chuang}.

\begin{definition}[Quantum density matrices]
    For a quantum system $A$ that can be in the state $\ket{\psi_j}$ with probability $p_j$, the quantum density matrix $\rho_A$ is defined as
    \begin{align}
        {\rho}_A = \sum_j p_j \ket{\psi_j}\bra{\psi_j},
    \end{align}
    with $p_j \geq 0$ and $\sum_j p_j =1$.
\end{definition}

\begin{definition}[Quantum operation]
    A quantum operation $\mathtt{E}$ from the quantum system $A$ to the quantum system $B$ is a linear, completely-positive, and trace-preserving map from the set of all density operators of the Hilbert space representing $A$ to the density operators of the Hilbert space representing $B$.
\end{definition} 

\begin{definition}[Trace and partial trace]
    Given a quantum operator $\mathtt{X}$ in a Hilbert space with dimension $d$, the trace operator is defined as
    \begin{align}
        tr(\mathtt{X}) = \sum_{i=0}^{d-1}\bra{i}\mathtt{X}\ket{i}.
    \end{align}
    In addition, given a composite quantum state $\rho_{AB}$, the partial trace for the quantum system $A$ is given by
    \begin{align}
        \rho_B = tr_{A}(\rho_{AB}) = \sum_{i=0}^{d_A-1} \left(\bra{i}\otimes I_{d_B}\right) \rho_{AB} \left(\ket{i} \otimes I_{d_B}\right),
    \end{align}
    where $d_A$ and $d_B$ are the dimensions of the Hilbert spaces associated with the quantum systems $A$ and $B$, respectively.
\end{definition}

\begin{theorem}[Kraus representation] \label{kraus}
    Any quantum operation $\mathtt{E}$ acting on a quantum state ${\rho}$ can be written in the form
    \begin{align}
        \mathtt{E}(\rho) = \sum_i {M}_i{\rho} {M}_i^{\dagger},
    \end{align}
    where $\sum_i {M}_i^\dagger {M}_i= {I}$.
\end{theorem}

\subsection{$N$-Sum Box Abstraction}
The $N$-sum box abstraction describes an approach for encoding the classical dits onto quantum states when the transmitters are entangled \cite{nsumbox}. The encoding and decoding structure in the $N$-sum box abstraction is described as follows. In the encoding stage, the servers use generalized Pauli operators $\mathtt{X}(a) = \sum_{j=0}^{q-1} \ket{j+a}\bra{j}$, and  $\mathtt{Z}(a) = \sum_{j=0}^{q-1} \omega^{tr(aj)} \ket{j}\bra{j}$, where $q=p^r$ with $p$ being a prime number, $a \in \mathbb{F}_q$ and $\omega = \exp(2\pi i /p)$, then transmit the resulting quantum states to the user over noise-free quantum channels. 

In the decoding stage, the user applies a projection-valued measurement (PVM) defined on the quotient space of the stabilizer group $\mathcal{L}(\mathcal{V})$ defined by
\begin{align}
    \mathcal{L}(\mathcal{V}) = \{c_{{v}} \mathtt{W}({v}) : {v} \in \mathcal{V} \},
\end{align}
where $\mathcal{V}$ is a self-orthogonal subspace, with respect to the symplectic inner product in $\mathbb{F}_q ^{2N}$ 
\begin{align}
     \mathtt{W}({v}) = \mathtt{X}(v_1) \mathtt{Z}(v_{N+1}) \otimes \ldots \otimes \mathtt{X}(v_N) \mathtt{Z}(v_{2N}),
\end{align}
and $c_{v} \in \mathbb{C}$ is chosen such that $\mathcal{L}(\mathcal{V})$ is an Abelian subgroup of $HW_{q}^N$  with $c_{{v}}{I}_{q^N}$ being an element of the stabilizer group only with $c_{v}=1$, where $HW_{q}^N$ is the Heisenberg-Weyl group defined as follows
\begin{align}
    HW_{q}^N = \{ c \mathtt{W}({s}) : {s} \in \mathbb{F}_q^{2N}, c \in \mathbb{C} \setminus \{0\} \}.
\end{align}

\begin{theorem}[Theorem~1  in \cite{nsumbox}]\label{feasible_nsum_1}
    Let ${G}$ be a $2N\times N$ SSO matrix, i.e., ${G^tJG}=0$, where ${J} = \begin{bmatrix}
        0 &{I}_N\\
        -{I}_N&0
    \end{bmatrix}$, and let ${H}$ be a $2N\times N$ matrix with full column rank. If $[{G} ~ {H}]$ is a full-rank matrix, then $M = [0_N ~ {I}_N ] [{G} ~ {H} ]^{-1}$ is a feasible $N$-sum box transfer matrix. 
\end{theorem}

\begin{figure*}[t]
    \centering
    \includegraphics[width=0.49\linewidth]{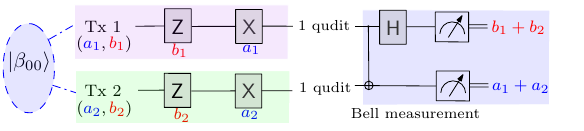}
    \hfill
    \includegraphics[width=0.49\linewidth]{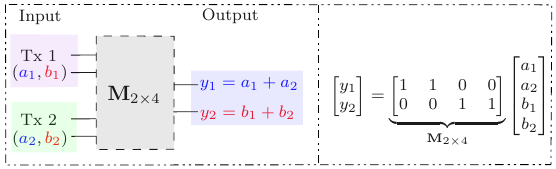}
    \caption{Quantum circuit for two entangled transmitters with the entangled state $\ket{\beta_{00}} = \frac{1}{\sqrt{2}} \left( \ket{00}+ \ket{11} \right)$ and the Bell measurement circuit on the left, and its $2$-sum box abstraction on the right using the transfer matrix $\mathbf{M}_{2 \times 4}$ to represent the input-output relationship, i.e., the entangled state and the Bell measurement, for the circuit on the left.}
    \label{fig:2_sum}
    \vspace*{0.4cm}
\end{figure*}

\begin{figure*}[t]
    \centering
    \includegraphics[width=0.49\linewidth]{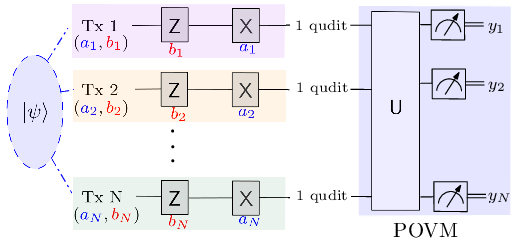}
    \hfill
    \includegraphics[width=0.49\linewidth]{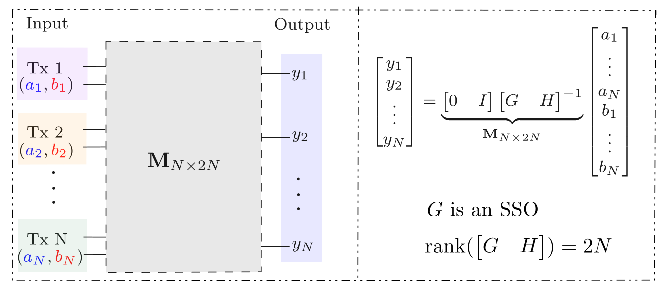}
    \caption{The general setting: Quantum circuit for $N$ entangled transmitters with the entangled state $\ket{\psi}$ with the corresponding quantum measurement circuit at the receiver side on the left, and its $N$-sum box abstraction on the right, using the transfer matrix $\mathbf{M}_{N \times 2N}$ to represent the input-output relationship, i.e., the entangled state and the measurement, for the circuit on the left.}
    \label{fig:n_sum}
\end{figure*}

A pictorial representation of the quantum circuit and its corresponding abstraction are given in Fig.~\ref{fig:2_sum} and Fig.~\ref{fig:n_sum}. In Fig.~\ref{fig:2_sum}, on the left side, the two transmitters are entangled using the pure state $\ket{\beta_{00}} = \frac{1}{\sqrt{2}} \left( \ket{00}+ \ket{11} \right)$, and in addition, the receiver applies the Bell measurement on the two received qubits. On the right side of the same figure, an abstract matrix $\mathbf{M}_{2 \times 4}$ is used to represent the relationship between the transmitted symbols $a_1,a_2,b_1$, $b_2$ with the received symbols after the Bell measurements, i.e., $a_1+a_2$ and $b_1+b_2$. Similarly, in Fig.~\ref{fig:n_sum}, the left side  represents a generalization for the $2$-sum circuit with an arbitrary number of servers $N$ sharing an entangled state $\ket{\psi}$ and the receiver applying a general POVM measurement on the received $N$ qudits. The right side of the same figure shows the relation between the inputs from all the $N$ transmitters $(a_1, \ldots, a_N, b_1, \ldots, b_N)$, where the $i$th transmitter sends $(a_i,b_i)$, and the $N$ received symbols after measurement on the receiver side using the transfer matrix $\mathbf{M}_{N \times 2N}$. The choice of a specific state $\ket{\psi}$ and POVM measurements corresponds to a specific transfer matrix $\mathbf{M}_{N \times 2N}$. Conversely, the choice of a specific value of $\mathbf{M}_{N \times 2N}$ satisfying the required conditions in Theorem~\ref{feasible_nsum_1} corresponds to one particular choice of $\ket{\psi}$ and POVM.

\section{Problem Formulation}\label{sec_prob_form}
We now formulate the PDMM problem in the classical and quantum setting, respectively.

In the classical setting, a user possessing two matrices $A$ and $B$, with constrained local computational resources, seeks to compute their product. These matrices are divided into $K$ and $L$ sub-blocks, respectively, as follows
\begin{align}
    A= \begin{bmatrix}
        A_1\\
        A_2\\
        \vdots\\
        A_K
    \end{bmatrix},  \qquad  B = \begin{bmatrix}
        B_1, B_2, \ldots, B_L
    \end{bmatrix},
\end{align}
where all $A_iB_j$ are compatible for computing the outer product, i.e., $AB$.
Given $K$ and $L$, the user distributes the encoded versions of the matrices to $N$ servers, such that, these matrices are secure against any $T$ colluding servers, i.e.,
\begin{align}\label{privacy_req}
    I(f_{\mathcal{T}_1}, g_{\mathcal{T}_2};A,B)=0,
\end{align}
where $f_i$ and $g_i$ are the information sent by the user to the $i$th server about $A$ and $B$, respectively.\footnote{It is assumed that $A$ and $B$ are generated independently and uniform at random  from the fields $\mathbb{F}_q^{Ks\times \kappa}$, and $\mathbb{F}_q^{\kappa \times \ell L}$, respectively, where $s,~\kappa, ~K,~L \in \mathbb{N}$.} In addition, $\mathcal{T}_i \subset [N]$, with $|\mathcal{T}_i| \leq T$, $i \in [2]$.

\begin{remark}
    Throughout this paper, we focus on the outer product partitioning case. In this case, we want to compute 
    \begin{align}\label{mult}
        AB = \begin{bmatrix}
            A_1B_1 & A_1B_2& \ldots & A_1B_L\\
           A_2B_1& A_2B_2 & \ldots& A_2B_L\\
           \vdots & \vdots & \ddots & \vdots\\
           A_K B_1 & A_KB_2& \ldots & A_KB_L
        \end{bmatrix}.
    \end{align}
\end{remark}

In order to ensure security for each of these matrices, a subspace with dimension at least $T$ has to be designed to protect the transmitted matrices. For the outer product, $AB$, this will result in a noise space with high dimensionality. Without careful design, the resulting code may suffer from rate inefficiency.

To address this, we adopt polynomial codes: Polynomial codes are considered as a special case of linear codes due to their nice structure that can be utilized in the analysis. A polynomial code for product computation is defined by two polynomials, $f: \mathbb{F}_q \rightarrow \mathbb{F}_q ^ {s\times \kappa }$ and $g:\mathbb{F}_q \rightarrow \mathbb{F}_q ^{\kappa \times \ell}$, and the goal is to compute $h=fg$ using the $N$ servers. Classically, once the user receives the responses, $h_{[N]}$, the required multiplication can be reconstructed, i.e., 
\begin{align}
    H(AB|h_{[N]})=0.
\end{align}
In addition, the rate can be defined as
\begin{align}
    R_{C}(K,L,T) = \frac{KL}{N}.
\end{align}

\begin{remark}
    It is important to note that, since only one polynomial, $f$, is used to encode the sub-blocks of $A_i$, sub-blocks have to be of equal dimensions, or zero padding must be used. The same argument holds for the sub-blocks of $B$.  {In addition, the optimization parameter is the number of servers $N$, i.e., fewer servers means fewer received bits since the operation is the same for any number of servers. }
\end{remark}

In the quantum setting, the servers share an entangled state in the quantum system $\mathtt{Q}^{\text{init}} = \mathtt{Q}_1^{\text{init}} \cdot \mathtt{Q}_2^{\text{init}} \cdot \ldots \cdot \mathtt{Q}_N^{\text{init}} $, where the $\mathtt{Q}^{\text{init}}_n$ subsystem is at a state represented by the density matrix $\rho^{\text{init}}_n = tr_{j \in [N] \setminus \{n\}}(\rho^{\text{init}})$, where $tr_{m}(\cdot)$ is the partial trace operator in the finite Hilbert space defined in Section \ref{review}. The user sends the sub-blocks' encoded versions, $f_n$ and $g_n$, to the $n$th server over a classical noiseless channel analogous to the classical PDMM protocol. Once the $n$th server receives the encoded versions, it applies classical operations and encodes over its own quantum state, $\rho_n^{\text{init}}$, to produce another quantum state $\rho_n$, 
\begin{align}
    \rho_n = \mathrm{Enc}(\rho_n^{\text{init}}, f_n,g_n).
\end{align}
Then, each server transmits $\rho_n$ to the user over a noise-free quantum channel. Upon receiving all quantum subsystems, the user applies an appropriate measurement to recover the classical output, $Y$, from the quantum states to decode the required multiplication, $AB$, i.e.,
\begin{align}\label{decod_req}
    H(AB|Y) = 0.
\end{align}

\begin{remark}
    Our encoding and decoding approaches follow the N-sum-box framework described in the subsequent sections.
\end{remark}

The efficiency, i.e., rate, for any scheme satisfying \eqref{privacy_req} and \eqref{decod_req} is defined as the ratio between the number of computations and the number of downloaded symbols, i.e., 
\begin{align}
    R(K,L,T) = \frac{mKL}{N},
\end{align}
where $m$ is an integer representing the number of instances for $AB$ that can be computed, {i.e., how many multiplications of the form in \eqref{mult} is obtained in each time the scheme is applied}. This is a generalization for the rate formulation; however, in the classical setting, we have $m=1$ and the classical rate is defined as $R_C(K,L,T)$. In the quantum setting, $m=2$, which will become clear in later sections. The quantum rate is denoted as $R_Q(K,L,T)$. 

\section{Classical and Quantum $\text{GASP}_r$ Codes}\label{sec_gasp_all}
In this section, we first review the construction of the classical GASP codes, along with some of their important properties. Then, we develop the quantum version of the GASP codes. Towards constructing the quantum GASP codes, we provide a feasibility condition that needs to be satisfied. Then, we provide the encoding and decoding structure for the quantum GASP codes once the aforementioned condition is satisfied. Afterwards, we investigate the feasibility condition analytically. We show that when $T \geq KL-K+1$, the GASP code can be extended to the quantum setting. 

\subsection{Classical $\text{GASP}_r$ Codes}
Now, we provide some conditions to ensure the decodability and privacy in polynomial codes. We then describe the classical GASP codes by defining the values of the exponents for each polynomial.

\begin{theorem}[Definition 5 and Theorem 1 in \cite{Gasp_1}] \label{decodability_def_classical}
    Let $\alpha \in \mathbb{N}^{K+T}$, $\beta \in \mathbb{N}^{L+T}$, the outer sum  $\alpha \oplus \beta$ code is decodable and $T$-private if 
    \begin{enumerate}
        \item $(\alpha \oplus \beta)_{k,\ell} \neq (\alpha \oplus \beta)_{k',\ell'}$, $\forall (k,\ell) \in [K]\times [L]$, and $(k',\ell') \in [K+T]\times [L+T]$, $(k,\ell) \neq (k',\ell')$, 
        \item $\alpha_{K+t} \neq \alpha_{K+t'}$ and $\beta_{L+t} \neq \beta_{L+t'}$, $\forall t \neq t'$.
    \end{enumerate}
\end{theorem}

\begin{definition}[\text{GASP} codes \cite{Gasp_1}]
    Given $K$, $L$ and $T$, for any $1 \leq r \leq \min\{K,T\}$, the $\text{GASP}_r$ code is defined by the following two polynomials $f(x) = \sum_{k=1}^K A_{k}x^{\alpha_k} + \sum_{t=1}^T R_{t}x^{\alpha_{K+t}}$, and $g(x) = \sum_{\ell=1}^L B_{\ell}x^{\beta_{\ell}} + \sum_{t=1}^T S_{t}x^{\beta_{L+t}} $, where $A_k$ and $B_{\ell}$ are the matrix partitionings, $R_t$ and $S_t$ are uniform noise random variables for masking, and the two polynomial exponents are chosen such that the decodability and $T$-privacy stated in Theorem \ref{decodability_def_classical} are satisfied as follows,
    \begin{align}
        \alpha_1 =& (0,1,\ldots, K-1),\\
        \alpha_2 =& (KL, KL+1, \ldots, KL+r-1, KL+K, \nonumber \\   
                  & ~  KL+K+1, \ldots, KL+K+r-1, \ldots)\text{ of size } T,\\
        \beta_1 =& (0,K, \ldots, K(L-1)),\\
        \beta_2 =& (KL, KL+1, \ldots, KL+T-1),\\
        \alpha =& (\alpha_1, \alpha_2),\\
        \beta =& (\beta_1, \beta_2),
    \end{align}
    with the number of servers $N$ satisfying $N = |Set(\alpha \oplus \beta)|$.
\end{definition}

\subsection{Generator Matrix Representation}
Let $\mathcal{S} = Set(\alpha \oplus \beta) \subset \mathbb{N}$ and fix an indexing for $\cs$ with $|\mathcal{S}| = N$. Then, the generator matrix for $\text{GASP}_r$ in a finite field $\mathbb{F}_{q^m}$, with $q \geq \cs(N)$, can be written as
\begin{align}
    G_{\text{GASP}_r} = \begin{bmatrix}
        a_1^{\cs}\\
        \vdots\\
        a_N^{\cs}
    \end{bmatrix},
\end{align}
where $a_1, a_2, \ldots, a_N$ are chosen uniformly at random from $\mathcal{G} \subseteq \mathbb{F}_{q^m}$,  $|\mathcal{G}| > \big(2{ N \choose T}+1\big)J $, with $J = \sum_{j \in \cs} j$, $a_i ^{\cs} = \begin{bmatrix} a_i^{\cs(1)}, \ldots, a_i^{\cs(N)} \end{bmatrix}$, and 
\begin{align}\label{main_opt}
    N =& KL+ 2K + 3T -2 - \max(K,\varphi) \nonumber \\  &+ (L-2)\max \left(0 , \min (r, r-\varphi) \right) \nonumber\\
    &+ \left\lfloor \frac{T-1}{r} \right\rfloor \min (T-1,K-r) \nonumber \\
    &- \mathbbm{1}_{\{\varphi < r\}} \cdot \bigg( \min(0, \mu-r) + r \frac{T-\mu-1}{K} \nonumber\\
    &+ \frac{Kx^2 +b x + T-\mu-1}{2}  - \frac{(T-\mu -1)(T+\mu -1)}{2K} \bigg), 
\end{align}
with $\varphi = T-1-KL+2K$, $\mu = (T-1) \ \mathrm{mod} \ K$, $x = \min\big(\frac{T-\mu-1}{K} - \mathbbm{1}_{\{\mu = 0\}}, L-3\big)$, and $b = -K-2\max(0,\varphi)+2T-2$.

\begin{remark}
    A detailed proof that the generator matrix $G$ is invertible with specific values of elements, which can be chosen prior to the scheme initiation, is provided in \cite[Appendix~A]{Gasp_1}.    
\end{remark}

\begin{remark}
    We note that the number of servers $N$ is also a function of $r$. However, we drop the function notation for simplicity if it is clear from the context.
\end{remark}

{To obtain the optimal value of the number of servers $N^*$ for a certain setting for given $K$, $L$ and $T$, the value of the free parameter $r$ is iterated over the specified range, i.e., $1 \leq r \leq \min\{K,T\}$. The value of $r$ that provides the least number of servers $N^*$ is denoted by $r^*$}
\subsection{Quantum $\text{GASP}_r$ Codes}
As in the previous section, we assume that the user shares a classical channel with the $N$ servers to send the encoded versions of the sub-matrices, $f_{[N]}$ and $g_{[N]}$. Upon receiving the encoded classical sub-matrices, the entangled servers apply classical and quantum operations individually, and transmit qudits back to the user over separate quantum channels. In this section, we provide the required condition such that the \text{GASP} codes can be extended to the quantum \text{GASP} codes and specify the corresponding scheme.

\subsubsection{Quantum $\text{GASP}_r$ Condition}
In this section, to specify the feasibility condition, we first define the longest consecutive chain (LCC) set-valued function and the partitioning of any matrix generated by $\alpha \oplus \beta$ into the information sub-space, upper left corner $(\mathcal{UL})$ in the case of GASP, and interference subspace $(\mathcal{IS})$.  

\begin{definition}
    Let LCC $: \mathcal{X} \rightarrow \mathcal{Y}$, where $\mathcal{X} \subset \mathbb{N}$, and $\mathcal{Y} \subseteq \mathcal{X}$. Then, LCC$(\cdot)$ returns the longest arithmetic progression with unit difference in $\mathcal{X}$.
\end{definition}

As a simple numerical example, let $\mathcal{X} = \{4,7,5,6,10,11,17\}$, then $LCC(\mathcal{X})= \{4,5,6,7\}$.

\begin{definition}
    Let ${PM}$ be a matrix in $\mathbb{N}^{K+T\times L+T}$, such that,
    \begin{align}
        {PM}(i,j) = \alpha(i) + \beta(j),
    \end{align}
    and define $\mathcal{UL}$ as the set 
    \begin{align}
        \mathcal{UL} =  Set\left(PM(1:K,1:L)\right).
    \end{align}
    The complement of $\mathcal{UL}$ is defined as $\mathcal{IS}$, where
    \begin{align}
        \mathcal{IS} =Set(PM) \setminus \mathcal{UL}. 
    \end{align}
\end{definition}

Now, to generate a scheme for the quantum GASP, we define the feasibility constraint mentioned in the previous sections. 

\begin{definition}[Feasibility constraint]\label{feasability_constraint}
    Let $r^*$ be the optimal parameter for \text{GASP}$(K,L,T)$ with the optimal number of servers $N^*_{\text{\text{GASP}}}(K,L,T)$. Define ${PM}$ as above with $\alpha$ and $\beta$ based on $r^*$. Then, the feasibility constraint is defined as    
    \begin{align}\label{eq_feas}
        |LCC(\mathcal{IS})| \geq  \left\lceil \frac{N^*_{\text{GASP}}(K,L,T)}{2}\right\rceil. 
    \end{align}
\end{definition}

\begin{theorem}[Quantum $\text{GASP}_r$]
    Given $K$, $L$ and $T$, let $N^*_{\text{\text{GASP}}}(K,L,T)$ be the optimal number of servers for the \text{GASP} codes. If \eqref{eq_feas} is satisfied, then there exists a quantum coding scheme for PDMM such that the rate is
    \begin{align}
        R_Q(K,L,T) = \frac{2KL}{N^*_{\text{\text{GASP}}}(K,L,T)} = 2 R_C(K,L,T).
    \end{align}
\end{theorem}

\subsubsection{Quantum GASP Scheme}
We now present the details of the proposed quantum GASP scheme. Since there are two instances of the classical scheme that can be applied, the user wants to compute $A^{[1]}B^{[1]}$ and $A^{[2]}B^{[2]}$ where each of them have the same structure as \eqref{mult}. To that end, the user sends two instances of the two polynomials $f(x)$ and $g(x)$ defined as 
\begin{align}
    f^{[m]}(a_i) &= \sum_{k=1}^K A_{k}^{[m]}a_i^{\alpha_k} + \sum_{t=1}^T R_{t}^{[m]}a_i^{\alpha_{K+t}},\\ 
    g^{[m]}(a_i) &= \sum_{\ell=1}^L B_{\ell}^{[m]}a_i^{\beta_{\ell}} + \sum_{t=1}^T S_{t}^{[m]}a_i^{\beta_{L+t}},
\end{align}
for $m \in [2]$, to each of the $N$ servers. Upon receiving them, each server applies the multiplication and computes 
\begin{align}\label{first_way}
    F^{[m]}(a_n) =& \sum_{k \in [K],~ \ell \in [L]} A_k^{{[m]}} B_{\ell}^{[m]} a_n^{(\alpha \oplus \beta)_{k,\ell}} \nonumber \\ &+ \sum_{(i,j) \in \mathcal{Q}} I_{i,j}^{[m]} a_n^{(\alpha \oplus \beta)_{i,j}}, ~ m \in [2],
\end{align}
where $\cq = \{(i,j) :~ i \in[K+T], ~ j \in [L+T], ~ (i,j) \notin ([K],[L])\}$, {and $I_{i,j}^{[m]}$ are interference terms that do not provide useful information for the required multiplication.}

Let $i_{k,\ell} = (\alpha \oplus \beta)_{k,\ell}$. Thus, for every $(k,\ell) \in [K]\times[L]$, $i_{k,\ell} \in \mathcal{UL}$. Similarly, for every $(i,j) \in \mathcal{Q}$, we represent $(\alpha\oplus \beta)_{i,j}$ as $c_{i,j}$ and $c_{i,j} \in \mathcal{IS}$. Thus, \eqref{first_way} can be written as
\begin{align}
    F^{[m]}(a_n) =& \sum_{i_{k,\ell} \in \mathcal{UL}} A_k^{{[m]}} B_{\ell}^{[m]} a_n^{i_{k,\ell}} + \sum_{c_{i,j} \in \mathcal{IS}} I_{c_{i,j}}^{[m]} a_n^{c_{i,j}}\\
    =& \sum_{i_{k,\ell} \in \mathcal{UL}} A_k^{{[m]}} B_{\ell}^{[m]} a_n^{i_{k,\ell}} + \sum_{c_{i,j} \in \mathcal{IS}\setminus \text{LCC}(\mathcal{IS})} I_{c_{i,j}}^{[m]} a_n^{c_{i,j}} \nonumber \\ &+ \sum_{c_{i,j} \in \text{LCC}(\mathcal{IS})} I_{c_{i,j}}^{[m]} a_n^{c_{i,j}}
    , ~ m \in [2]. \label{comp_1}
\end{align}
To make the notation easy, first we define $\mathcal{S} = LCC(\mathcal{IS})$, and $\mathcal{R} =  \mathcal{IS}\setminus LCC(\mathcal{IS})$. In addition, note that the running index pair $(i,j)$ for the interference terms can be substituted by $\cs$ and $\cR$ as follows
\begin{align}
    F^{[m]}(a_n) =&  \sum_{i_{k,\ell} \in \mathcal{UL}} A_k^{{[m]}} B_{\ell}^{[m]} a_n^{i_{k,\ell}} + \sum_{r_i \in \mathcal{R}:  \atop i \in [|\mathcal{R}|]} I_{r_i}^{[m]} a_n^{r_i} \nonumber \\ &+ \sum_{s_i \in \mathcal{S}:  \atop i \in [|\mathcal{S}|]} I_{s_i}^{[m]} a_n^{s_i}, ~ m \in [2].
\end{align}
Now, for each element in $F_n^{[1]} = F^{[1]}(a_n)$, $F_n^{[2]} = F^{[2]}(a_n)$, the server applies Pauli operators on each element as follows
\begin{align}
    U_n(i,j) \rho_n
    U_n(i,j)^{\dagger},
\end{align}
where $U_n(i,j) =\mathtt{X}(u_n F_n^{[1]}(i,j))\mathtt{Z}(v_n F_n^{[2]}(i,j))$. 
Let us denote the optimal number of servers as $N^*$ for ease of notation. Thus, the transmitted symbols for each $(i,j)$ is given by
\begin{align}
   X(i,j) \!\!=\!\!& \begin{bmatrix}
        F_1^{[1]}(i,j)\\
        \vdots \\
        F_{N^*}^{[1]}(i,j) \\
        F_1^{[2]}(i,j)\\
        \vdots\\
        F_{N^*}^{[2]}(i,j)       
        \end{bmatrix} \\
        =\!\! & \begin{bmatrix}
        \diag(u)G_{\text{GASP}_{N^*}} & 0_{N^*}\\
        0_{N^*} & \diag(v)G_{\text{GASP}_{N^*}}
        \end{bmatrix} \nonumber \\ 
        &\times \Big[ (A_1B_1)^{[1]}(i,j),
        \ldots,
        (A_KB_L)^{[1]} (i,j),\nonumber\\
        & \!\!\qquad
        I_1^{[1]}(i,j),
        \ldots,
        I_{N^* - KL}^{[1]}(i,j),(A_1B_1)^{[2]}(i,j),
        \ldots, \nonumber\\
        & \!\!\qquad (A_KB_L)^{[2]} (i,j),
        I_1^{[2]}(i,j),
        \ldots,
        I_{N^* - KL}^{[2]}(i,j) \Big]^t
        \\
        =& D \begin{bmatrix}
        Q& 0\\
        0& Q
        \end{bmatrix} \nonumber\\
        &\times \Big[ I_{s_1}^{[1]}(i,j),
        I_{s_2}^{[1]}(i,j),
        \ldots,
        I_{s_{|\mathcal{S}|}}^{[1]}(i,j), 
        (A_1B_1)^{[1]}(i,j),\nonumber\\
        & \qquad
        \ldots,
        (A_KB_L)^{[1]} (i,j),  I_{r_1}^{[1]}(i,j),
        \ldots,
         \nonumber\\
        & \qquad I_{r_{|\mathcal{R}|}}^{[1]}(i,j),
        I_{s_1}^{[2]}(i,j),
        I_{s_2}^{[2]}(i,j),
        \ldots, \nonumber\\
        & \qquad 
        I_{s_{|\mathcal{S}|}}^{[2]}(i,j), (A_1B_1)^{[2]}(i,j), \ldots,
        (A_KB_L)^{[2]} (i,j), \nonumber\\
        &  \qquad      I_{r_1}^{[2]}(i,j),
        \ldots,
        I_{r_{|\mathcal{R}|}}^{[2]}(i,j) \Big]^t,
\end{align}
where $D=\begin{bmatrix} \diag(u)& 0_{N^*}\\ 0_{N^*} & \diag(v) \end{bmatrix}$, and
$Q = \begin{bmatrix} Q_1|Q_2 \end{bmatrix}$, where
\begin{align}
    Q_1 &= \begin{bmatrix}
    a_1^{s}& a_1^{s+1} & \ldots &  a_1^{s+|\mathcal{S}|-1}\\
    a_2^{s}& a_2^{s+1} & \ldots &  a_2^{s+|\mathcal{S}|-1}\\
    \vdots & \vdots & \ddots &  \vdots \\
    a_{N^*}^{s}& a_{N^*}^{s+1} & \ldots &  a_{N^*}^{s+|\mathcal{S}|-1}\\
    \end{bmatrix},\\
    Q_2 &= \begin{bmatrix}
       a_1^{i_{1,1}}& a_1^{i_{1,2}}& \ldots & a_1^{i_{K,L}} & a_1^{r_1} & \ldots & a_1^{r_{|\mathcal{R}|}}\\
         a_2^{i_{1,1}}& a_2^{i_{1,2}}& \ldots & a_2^{i_{K,L}} & a_2^{r_1} & \ldots & a_2^{r_{|\mathcal{R}|}}\\
         \vdots & \ddots & \vdots & \vdots & \ddots & \vdots\\
         a_{N^*}^{i_{1,1}}& a_{N^*}^{i_{1,2}}& \ldots & a_{N^*}^{i_{K,L}} & a_{N^*}^{r_1} & \ldots & a_{N^*}^{r_{|\mathcal{R}|}}
    \end{bmatrix},
\end{align}

We next design a transfer matrix $M$ that satisfies the requirements in Theorem \ref{feasible_nsum_1}. Accordingly, we construct the following matrix and show that it satisfies the required properties
\begin{align}
    M = \begin{bmatrix}
        0& I_{N^*}
        \end{bmatrix}
        \begin{bmatrix}
        G & H
        \end{bmatrix}^{-1},
\end{align}
where 
\begin{align}
    G = \begin{bmatrix}
        DQ(:,1:\lfloor \frac{N^*}{2}\rfloor) & 0\\
        0 & DQ(:,1:\lceil \frac{N^*}{2}\rceil)
        \end{bmatrix},
\end{align}
and 
\begin{align}
    H = \begin{bmatrix}
        DQ(:,\lfloor\frac{N^*}{2}\rfloor + 1: N^*) & 0 \\
        0& DQ(:,\lceil\frac{N^*}{2}\rceil + 1: N^*)
    \end{bmatrix}.
\end{align}
Consequently, the received symbols are given by 
\begin{align}
    Y(i,j) =& M X(i,j)\nonumber\\ =& 
    \Big[
    I_{s_{\lfloor\frac{N}{2}\rfloor +1}}^{[1]}(i,j),
    \ldots,
    I_{s_{|\mathcal{S}|}}^{[1]}(i,j),
    (A_1B_1)^{[1]}(i,j),\nonumber \\
    &
    \ldots,
    (A_KB_L)^{[1]} (i,j),  I_{r_1}^{[1]}(i,j),
    \ldots,
    I_{r_{|\mathcal{R}|}}^{[1]}(i,j),\nonumber \\
    &
    I_{s_{\lceil\frac{N}{2}\rceil +1}}^{[2]}(i,j), 
    \ldots,
    I_{s_{|\mathcal{S}|}}^{[2]}(i,j), (A_1B_1)^{[2]}(i,j),
    \ldots, \nonumber\\
    &
    (A_KB_L)^{[2]} (i,j),
    I_{r_1}^{[2]}(i,j),
    \ldots,
    I_{r_{|\mathcal{R}|}}^{[2]}(i,j)
    \Big]^t.
\end{align}
As shown, the scheme provides two instances in each $N^*$ qubits transmission compared to one instance for the classical counterpart. Thus, the rate is given by $R_Q = 2R_C$.

Now, we need to show that the chosen $\begin{bmatrix} G& H \end{bmatrix}$ satisfies the requirements in Theorem \ref{feasible_nsum_1}. First, it follows directly that 
\begin{align}
    \begin{bmatrix}
        \diag(u)G_{\text{GASP}_{N^*}} & 0_{N^*}\\
    0_{N^*} & \diag(v)G_{\text{GASP}_{N^*}}
    \end{bmatrix}
\end{align}
is invertible if $G_{\text{GASP}}$ is invertible. Since applying any column permutation will not affect the span of the column space, the required matrix is invertible. 

Finally, we  verify that $G$ is an SSO matrix,
\begin{align}
    G = &\begin{bmatrix}
        DQ(:,1:\lfloor \frac{N^*}{2}\rfloor) & 0\\
        0 & DQ(:,1:\lceil \frac{N^*}{2}\rceil)
    \end{bmatrix}\\
    =& \begin{bmatrix}
    \diag(u)& 0_{N^*}\\
    0_{N^*} & \diag(v)
    \end{bmatrix} \nonumber \\
    &\times\begin{bmatrix}
    a_1^{s}\!&\!a_1^{s+1}\!&\!\ldots\!&\!a_1^{s+\mu_1}\!&\!0\!&\!0\!&\!\ldots\!&\!0\\
    a_2^{s}\!&\!a_2^{s+1}\!&\!\ldots\!&\!a_2^{s+\mu_1}\!&\!0\!&\!0\!&\!\ldots\!&\!0\\
    \vdots\!&\!\vdots\!&\!\ddots\!&\!\vdots\!&\!\vdots\!&\!\vdots\!&\!\ddots\!&\!\vdots\\
    a_{N^*}^{s}\!&\!a_{N^*}^{s+1}\!&\!\ldots\!&\!a_{N^*}^{s+\mu_1}\!&\!0\!&\!0\!&\!\ldots\!&\!0\\
    0\!&\!0\!&\!\ldots\!&\!0\!&\!a_1^{s}\!&\!a_1^{s+1}\!&\!\ldots\!&\!a_1^{s+\mu_2}\\
    0\!&\!0\!&\!\ldots\!&\!0\!&\!a_2^{s}\!&\!a_2^{s+1}\!&\!\ldots\!&\!a_2^{s+\mu_2}\\
    \vdots\!&\!\vdots\!&\!\ddots\!&\!\vdots\!&\!\vdots\!&\!\vdots\!&\!\ddots\!&\!\vdots\\
    0\!&\!0\!&\!\ldots\!&\!0\!&\!a_{N^*}^{s}\!&\!a_{N^*}^{s+1}\!&\!\ldots\!&\!a_{N^*}^{s+\mu_2}
    \end{bmatrix},
\end{align}
where $\mu_1 = \lfloor \frac{N^*}{2}\rfloor -1$, and $\mu_2 = \lceil \frac{N^*}{2}\rceil -1$.
To prove that $G$ satisfies the SSO condition if $u$ and $v$ are chosen to generate dual codes, we provide the following theorem.

\begin{theorem}
    Let $G = \begin{bmatrix}
        M_{n\times k} & 0_{ n\times n-k}\\
        0_{n \times k} & S_{n \times n-k}
    \end{bmatrix}$. If $M$ and $S$ are generators of two dual codes, then $G$ satisfies the SSO condition.
\end{theorem}
The proof of the theorem is shown by just computing the value of $G^tJG$ in terms of $M$ and $S$, recalling that they are dual. Thus, $G^tJG = 0$ is satisfied when the values of $u$ and $v$ are chosen so that the shifted GRS codes are dual. 

\subsection{Applicability of the Feasibility Condition}
In this section, we provide an analytical result for the feasibility condition. We prove that if $T \geq KL - K +1$, for non-degenerate cases, i.e., $K,~L \geq 2$, then the feasibility condition is always satisfied. We formulate this in the following theorem.

\begin{theorem}\label{thm_feas_analysis}
    If $T \geq KL-K+1$ and $K, L \geq 2$, then the condition $ |LCC(\mathcal{IS})| \geq  \left\lceil \frac{N^*_{\text{GASP}}(K,L,T)}{2}\right\rceil$ in \eqref{eq_feas} is satisfied and the GASP code can be extended to the quantum counterpart.
\end{theorem}

\begin{Proof}
    We start by looking into all the terms in \eqref{main_opt}. Then, we recall that
    \begin{align}
        \varphi = T-1-KL+2K  \geq K.
    \end{align}
    In addition, we have $r \leq K$, implying $r \leq \varphi$. This allows us to remove the last term in \eqref{main_opt} that is multiplied by $\mathbbm{1}_{\{r > \varphi\}} = 0$. Note that $\min(r,r-\varphi) = r-\varphi$ and $\max(0, r-\varphi) = 0$. This reduces \eqref{main_opt} to 
    \begin{align}
        N =& KL+2K + 3T-2-\max(K,\varphi) \nonumber \\ &+ \left\lfloor\frac{T-1}{r}\right\rfloor \min(T-1,K-r) \\
        =& KL+2K + 3T-2-\varphi\nonumber \\ &+ \left\lfloor\frac{T-1}{r}\right\rfloor \min(T-1,K-r)\\
        =& 2KL+2T-1 + \left\lfloor\frac{T-1}{r}\right\rfloor \min(T-1,K-r).
    \end{align}
    Note that $\lfloor\frac{T-1}{r}\rfloor \geq 0$, and $\min(T-1,K-r) \geq 0$, and thus
    \begin{align}
     N^* = 2KL+2T-1,   
    \end{align}
    which is achievable when $r = K$. Now, we need to focus on the interference subspace. We note that the exponents when $r=K$ are given by
    \begin{align}
         \alpha_1 =& (0,1,\ldots, K-1),\\
        \alpha_2 =& (KL, KL+1, \ldots, KL+T-1),\\
        \beta_1 =& (0,K, \ldots, K(L-1)),\\
        \beta_2 =& (KL, KL+1, \ldots, KL+T-1).
    \end{align}
    The interference subspace lies in $\alpha_1\oplus \beta_2$, $\beta_2\oplus \alpha_1$ and $\alpha_2 \oplus \beta_2$. Note that $Set(\alpha_1 \oplus \beta_2) = [KL:KL+T+K-2]$. In addition, $Set(\alpha_2 \oplus \beta_2) = [2KL:2KL+2T-2]$. For the third interval, we have
    \begin{align}
        Set(\alpha_2\oplus\beta_1) = \bigcup_{j=0}^{L-1} [KL+jK : KL+jK+T-1].
    \end{align}
     Then, to check how these intervals overlap, we check $KL+jK+T-1-(KL+(j+1)K) = -K+T-1\geq 0 $ since $T \geq KL-K+1 $ with $L \geq 2$. This proves that
     \begin{align}
         Set(\alpha_2\oplus\beta_1) &= \bigcup_{j=0}^{L-1} [KL+jK : KL+jK+T-1] \\
         &= [KL:KL+(L-1)K+T-1]\\
         &=[KL:2KL+T-K-1].
     \end{align}
     Now, to check the relation between $Set(\alpha_1\oplus\beta_2)$ and $Set(\alpha_2\oplus\beta_1)$, we note again that $L\geq 2$, and thus $2KL+T-K-1 \geq KL+T+K-2$. Hence,
     \begin{align}
         Set(\alpha_1\oplus\beta_2) \cup Set(\alpha_2\oplus\beta_1) = [KL:2KL+T-K-1].
     \end{align}
     Finally, since $2KL+T-K-1 \geq 2KL$ and $2KL+2T-2 \geq 2KL +T-K-1$, the interference set is given by
     \begin{align}
         \mathcal{IS} = [KL:2KL+2T-2].
     \end{align}
     Since this is a set with no gaps, we have $LCC(\mathcal{IS}) = \mathcal{IS} = [KL:2KL+2T-2]$ with cardinality $|LCC(\mathcal{IS})| = KL+2T-1$. Then, we need to check the feasibility condition \eqref{eq_feas}. Recall that $N^* = 2KL+2T-1$ which is odd, then $\lceil \frac{N^*}{2}\rceil = KL+T$. Finally, since $|LCC(\mathcal{IS})| = KL+2T-1 \geq \lceil \frac{N^*}{2}\rceil = KL+T$, the feasibility constraint is satisfied, concluding the proof.
\end{Proof}

\begin{remark}
    Theorem~\ref{thm_feas_analysis} implies that if $T \geq KL$ and $K, ~L \geq 2$, then the condition $ |LCC(\mathcal{IS})| \geq  \left\lceil \frac{N^*_{\text{GASP}}(K,L,T)}{2}\right\rceil$ in \eqref{eq_feas} is satisfied and the GASP code can be extended to the quantum counterpart.
\end{remark}

\begin{remark}
    Theorem~\ref{thm_feas_analysis} implies that we only need to focus on designing new quantum codes for the case when $T < KL-K+1$. 
\end{remark}

\subsection{Discussion and Further Remarks}
\begin{figure}[t]
    \centering
    \includegraphics[width=\linewidth]{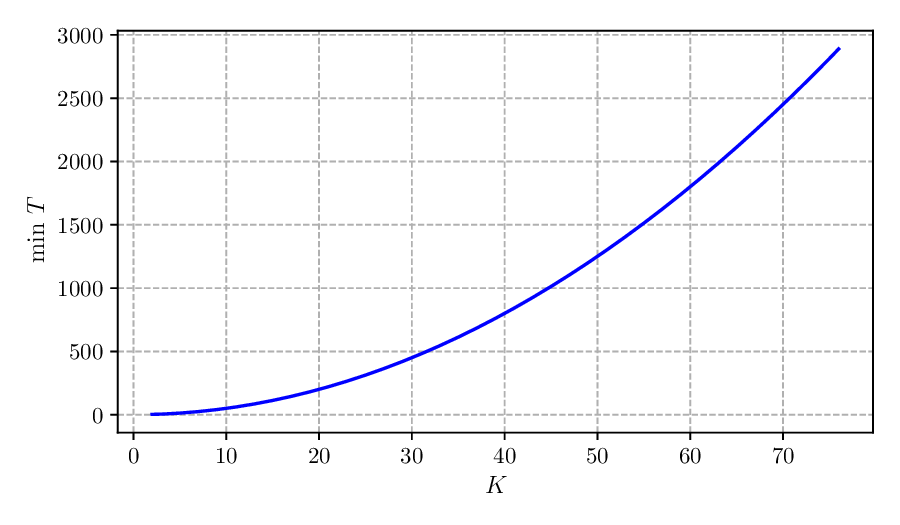}
    \caption{Min $T$ needed for $K=L$ to satisfy \eqref{eq_feas}.}
    \label{Gasp_K_eq_L}
\end{figure}

Although Theorem \ref{thm_feas_analysis} ensures that when $T \geq KL -K+1$ the feasibility condition is satisfied, it does not rule out the case when $T < KL-K+1$. The issue with further analytical investigation is that the condition $\varphi \geq r$ for all $r$ is not satisfied anymore. Thus, in order to understand roughly and intuitively the relationship between the feasibility condition and the GASP code, we analyze it numerically. To quantify the applicability of the feasibility condition \eqref{eq_feas}, we provide an estimate for the least value of the privacy parameter needed in order to extend the classical GASP codes to the quantum GASP. Fig.~\ref{Gasp_K_eq_L} shows that in the case of $K=L$, as $K$ increases, the minimum value of $T$, $\hat{T}_{\rm min}$, required to satisfy the feasibility constraint increases quadratically in the following way to the nearest integer
\begin{align}
    \Hat{T}_{\rm min} = 0.5 K^2 - 10^{-3}K + 0.772.
\end{align}

\begin{figure}[t]
    \centering
    \includegraphics[width=\linewidth]{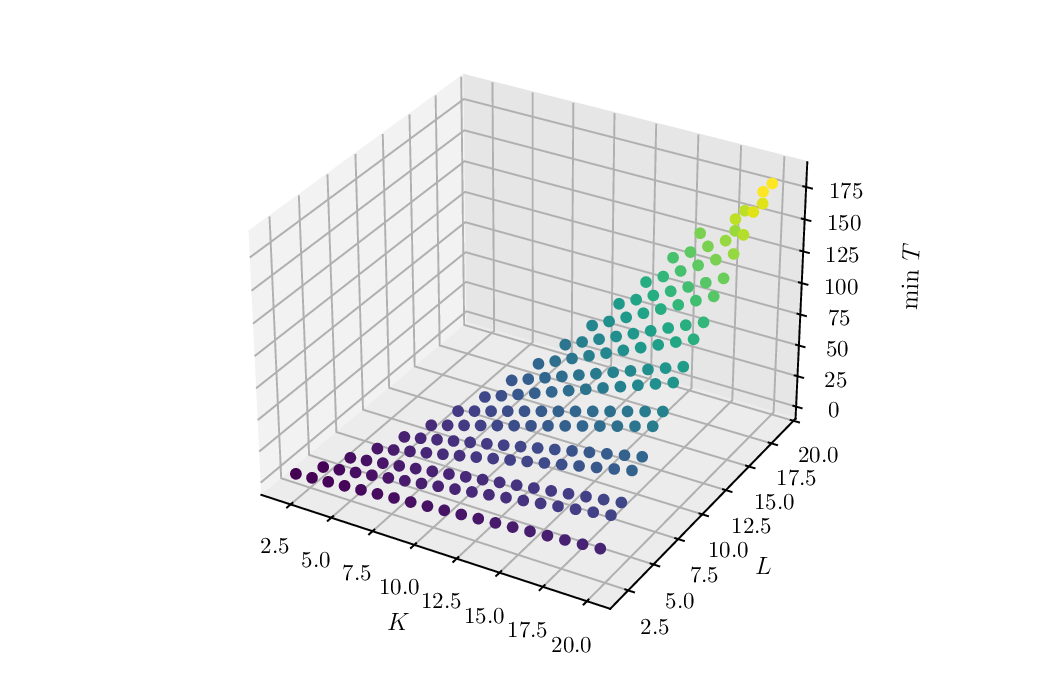}
    \caption{Min $T$ needed for $K$ and $L$ to satisfy \eqref{eq_feas}.}
    \label{Gasp_K_neq_L}
\end{figure}

To move from the $K=L$ case to the $K \geq L$ case, Fig.~\ref{Gasp_K_neq_L} shows the $3D$ plot relating the three variables. It is estimated, using multi-variable regression up to quadratic terms, as 
\begin{align}\label{t_min_1_2_eq}
    \Hat{T} _{\rm min} =  -0.043 L^2 + 0.507KL + 0.18K + 0.362L  -0.746.
\end{align}

The $R^2$ score (which explains the percentage of variation in the output that the model can explain) in this case is $0.998$. $R^2$ ranges from  $- \infty$ to $1$, with the typical range being between $0$ and $1$, with $1$ being the perfect fit and $0$ meaning it is a horizontal line exactly in the midpoint of the samples. Even though the $R^2$ score is high and the estimate of the case when $K=L$ matches well with the real curve, still these relations should be interpreted as empirical approximations based on the specified parameter range. 

Two potential research directions emerge from the results presented in this section. The first is to investigate alternative methods for constructing dual codes for the GASP, that relax the feasibility requirement in \eqref{eq_feas}. The other direction is to consider all possible $r$ given $K$, $L$ and $T$. This might give more possibilities for satisfying the feasibility constraint in \eqref{eq_feas} with acceptable sub-optimality.

From the $\Hat{T}_{\rm min}$ estimate relation for $K \geq L$ in \eqref{t_min_1_2_eq}, we see that $T$ is affected more by varying $L$ than $K$. Consequently, if $L$ can be maintained to be small, then it is possible to easily satisfy the feasibility constraint \eqref{eq_feas}. 

\section{PDMM Codes for Low Privacy Regimes}\label{sec_more_pdmm_low_privacy}
Prior to delving into a newly discovered PDMM codes, the following definition, adopted from \cite{cat-dog}, will be useful for most of the defined codes.

\begin{definition}
    Let $gap(\ell,x,r) \in \mathbb{Z}^{\ell}$ be the generalized progression of length $\ell$ and chain length $r$, such that
    \begin{align}
        gap(\ell,x,r) = [&0,1,2, \ldots ,r-1, \nonumber \\ 
        & x, x+1,x+2, \ldots, x+r-1, \nonumber \\ 
        & 2x, 2x+1, 2x+2, \ldots, 2x+r-1 , \ldots].
    \end{align}
\end{definition}

The next three coding schemes only provide better rates than the GASP$_r$ when $T \leq K$, except for optimized GASP, which consistently does not perform worse than GASP. However, it is challenging to come up with situations where they provide better rates and can satisfy \eqref{eq_feas}. This is mainly because, in order to provide more efficiency to the schemes, more gaps in the exponents are added to decrease the number of servers. However, in one case for the CAT$_x$ code, we are able to satisfy the required constraint \eqref{eq_feas} when $T=K=L = 2$, and have better efficiency than GASP. This shows that there can exist cases for these three codes that can be extended to their quantum version, and at the same time, have better efficiency than the GASP$_r$ code.

To that end, we rewrite the construction of the three aforementioned codes and design the quantum CAT$_x$ code for the special case of $K=L=T=2$. We note again that these codes can possibly be extended to the quantum case if \eqref{eq_feas} is satisfied.

\begin{definition}[Optimized GASP (GASP$_{r,s}$) codes]
    Given $K$, $L$ and $T$ $\geq 2$, let $1 \leq r,s \leq \min(K,T)$. Then, the GASP$_{r,s}$ exponents are given by
    \begin{align}
        \alpha_1 &= [0:K-1], \\
        \alpha_2 &= KL+ \text{gap}(T,K,r),\\
        \beta_1 &= K \times [0:L-1], \\
        \beta_2 &= KL + \text{gap}(T,K,s).
    \end{align}
\end{definition}

\begin{definition}[DOG$_{r,s}$ codes]
    Given $K$, $L$ and $T$ $\geq 2$, let $1 \leq r \leq T$ and $1 \leq s \leq \min(K+r,T)$. Then, the DOG$_{r,s}$ exponents are given by,
    \begin{align}
        \alpha_1 &= [0:K-1], \\
        \alpha_2 &= K+ \text{gap}(T,K+r,r),\\
        \beta_1 &= (K+r)\times [0:L-1], \\
        \beta_2 &= (K+r)(L-1)+K + \text{gap}(T,K+r,s).
    \end{align}
\end{definition}

\begin{definition}[CAT$_x$ codes]
    For $K\geq L \geq T \geq 2$, let $\kappa$ and $\lambda$ be the smallest non-negative integers that make $K^* = K+1+\kappa$ and $L^* = L+1+ \lambda$ be co-primes with $\Bar{T} = T-1$. Let $q=K^*L^*+\Bar{T}^2$, and $x$ be any number co-prime with $q$. Finally, let $y$ be the solution to $x\Bar{T} + yK^* = 0 ~(mod ~q)$. Then, the exponents of CAT$_x(K,L,T)$ are given by
    \begin{align}
        \alpha_1 &= y\times[0:K-1] ~mod~ q,\\
        \alpha_2 &= x\times[0:T-1]+Ky ~mod~ q,\\
        \beta_1 &= x\times[0:L-1] ~mod~ q,\\
        \beta_2 &= y\times[0:T-1]-x ~mod~ q.
    \end{align}
\end{definition}

\subsection{Schr\"{o}dinger CAT$_x(2,2,2)$ Code}
Instead of calling it the quantum CAT code, we choose that name for obvious reasons. In this section, we extend the classical CAT$_x(2,2,2)$ to the quantum version since this is the only code with these specific parameters that we are able to find that provides a better rate than GASP$_r$ and at the same time satisfies \eqref{eq_feas}. 

Prior to delving into the quantum encoding, we refer the reader to the classical construction of this specific example in \cite{cat-dog}. To summarize, $N=q=10$, the root of unity chosen is $2$ and $\omega_i = 2^{i-1}$. Thus, the generator matrix can be written as $[n,n,d]_{11}$ GRS$(\omega,u) = \text{GRS}_{\omega,u}$, where $u = [1,1,\ldots,1]$. We can write the classical transmitted symbols as
\begin{align}
    Y(i,j) =& \text{GRS}_{\omega,u} 
       \Big[ A_1B_1(i,j),
        A_1B_2(i,j), 
        I_1(i,j),
        A_2B_1(i,j),  \nonumber \\
        & \qquad \qquad
        A_2B_2(i,j), I_2(i,j),
        I_3(i,j), 
        I_4(i,j),  \nonumber \\
        & \qquad \qquad
        I_5(i,j), 
        I_6(i,j) \Big]^t\\
        =& \begin{bmatrix}
          G_1 & H_1
        \end{bmatrix} 
        \Big[ I_2(i,j),
        I_3(i,j), 
        I_4(i,j), 
        I_5(i,j), 
        I_6(i,j), \nonumber\\
        &\qquad \qquad \quad A_1B_1(i,j),
        A_1B_2(i,j), 
        I_1(i,j),\nonumber\\
        &\qquad \qquad \quad
        A_2B_1(i,j),
        A_2B_2(i,j){\Big]}^t,
\end{align}
where $G_1 = \text{GRS}_{\omega,u}(:,6:10)$ and $H_1 = \text{GRS}_{\omega,u}(:,1:5)$. 

It is straightforward to verify that $G_1$ can be considered as $[10,5,d_1]_{11}$ $5$-shifted GRS code. The $5$-shifted $[10,5,d_2]_{11}$ dual code can be generated by $\omega$ and $v$, where $v$ satisfies \eqref{dual_grs_relation} with $\ell_1 = \ell_2 = 5$. Let the generator matrix of the dual code be defined as $G_2$, and it can be verified that $\begin{bmatrix} G_1& 0 \\ 0& G_2 \end{bmatrix}$ is an SSO matrix. In addition, we choose $H_2 = \text{GRS}_{\omega,v}(:,1:5)$. The received symbols from the quantum scheme can be written as
\begin{align}
    Y(i,j) =& \begin{bmatrix}
        0 & I
        \end{bmatrix} \begin{bmatrix}
        \!G_1 & \!0 & \!H_1 & \!0\\
        \!0 & \!G_2 & \!0 & \!H_2
        \end{bmatrix}^{-1} 
        \begin{bmatrix}
        \!G_1 & \!0 & \!H_1 & \!0\\
        \!0 & \!G_2 & \!0 & \!H_2
        \end{bmatrix} \nonumber \\     
        &\times \Big[ I_2^{[1]}(i,j),
        I_3^{[1]}(i,j), 
        I_4^{[1]}(i,j), 
        I_5^{[1]}(i,j), 
        I_6^{[1]}(i,j), \nonumber\\
        & \qquad I_2^{[2]}(i,j),
        I_3^{[2]}(i,j), 
        I_4^{[2]}(i,j), 
        I_5^{[2]}(i,j), 
        I_6^{[2]}(i,j), \nonumber\\
        & \qquad A_1B_1^{[1]}(i,j),
        A_1B_2^{[1]}(i,j), 
        I_1^{[1]}(i,j),\nonumber\\
        & \qquad
        A_2B_1^{[1]}(i,j),
        A_2B_2^{[1]}(i,j),  A_1B_1^{[2]}(i,j),\nonumber\\
        & \qquad
        A_1B_2^{[2]}(i,j), 
        I_1^{[2]}(i,j),
        A_2B_1^{[2]}(i,j),\nonumber\\
        & \qquad
        A_2B_2^{[2]}(i,j)\Big]^t \\
        =& \Big[  A_1B_1^{[1]}(i,j),
        A_1B_2^{[1]}(i,j), 
        I_1^{[1]}(i,j),
        A_2B_1^{[1]}(i,j),\nonumber\\
        &
        A_2B_2^{[1]}(i,j),   A_1B_1^{[2]}(i,j),
        A_1B_2^{[2]}(i,j), 
        I_1^{[2]}(i,j),\nonumber\\
        &
        A_2B_1^{[2]}(i,j),
        A_2B_2^{[2]}(i,j){\Big]}^t.
\end{align}
Thus, we are able to achieve a quantum rate that is double the highest classical state-of-the-art version, i.e., $R_Q = 4/5$.

\subsection{Discussion and Further Remarks}
Although we are not able to find many situations where the optimal code out of the three codes, optimized GASP, CAT and DOG, satisfy the feasibility constraint in \eqref{eq_feas} and at the same time have better performance than GASP, here we presented a simple example that shows that it indeed is possible. This makes the search for a relaxed version of the feasibility requirement an urgent matter so that these codes can be extended to their quantum counterparts.

\section{Conclusion}
In this paper, we investigated the problem of private distributed matrix multiplication (PDMM) for the case of outer product partitioning (OPP) in the quantum setting, i.e., the QPDMM problem. First, we formulated the quantum setting for the PDMM problem when the servers share an entangled state. Second, we divided our approach into two regimes based on the privacy requirement. In the first regime, we extended the state-of-the-art classical coding scheme, i.e., gap additive secure polynomial (GASP) code, when a feasibility condition is always satisfied. We showed that when this condition is satisfied, super-dense coding gain is achieved, i.e., the rate of the quantum setting is double its counterpart in the classical setting. We showed this by developing a quantum scheme that can encode two instances of the GASP code in the quantum setting instead of one instance in the classical setting. Third, we analyzed the feasibility condition to know when the GASP code can be extended to its quantum counterpart. We proved that when $T \geq KL-K+1$, the feasibility condition is satisfied. We then observed numerically that the minimum privacy required for the feasibility condition to be satisfied is of the order of $0.5KL$, however, we were not able to prove it analytically.
Finally, we investigated the low privacy regime in the same framework. We analyzed how to extend the state-of-the-art classical codes, i.e., the cyclic-addition degree table (CAT) and the discretely optimized GASP (DOG) codes. We gave a specific numerical example in which the CAT code can be extended to the quantum setting, achieving the super-dense coding gain. 

\bibliographystyle{unsrt}
\bibliography{references}
\end{document}